\documentclass[12pt,draftclsnofoot,onecolumn]{IEEEtran}
\ifCLASSINFOpdf
   \usepackage[pdftex]{graphicx}

\else

\fi

\usepackage[cmex10]{amsmath}
\usepackage{amsfonts}
\usepackage{amssymb}
\usepackage{subfig}
\usepackage{epstopdf}
\usepackage{algorithmic}
\usepackage{array}

\usepackage{mdwmath}
\usepackage{mdwtab}
\usepackage{citesort}
\usepackage{balance}
\usepackage{cite}
\usepackage{graphicx}
\begin{document}
%\setlength{\floatsep}{-1pt}
%\setlength{\intextsep}{-2pt}
\title{An Asymptotic Bound on Estimation and Prediction of Mobile MIMO-OFDM Wireless Channels}

\author{Ramoni~O.~Adeogun,~\IEEEmembership{Student~Member,~IEEE,}
        Paul~D.~Teal,~\IEEEmembership{Senior~Member,~IEEE,}
        and~Pawel~A.~Dmochowski,~\IEEEmembership{Senior~Member,~IEEE}% <-this % stops a space
\thanks{The authors are with the School of Engineering and Computer Science, Victoria University of Wellington, 
New Zealand (E-mail: [ramoni.adeogun, paul.teal, pawel.dmochowski]@ecs.vuw.ac.nz)}}% <-this % stops a space

%\author{Author~1, Author~2, Author~3}

%\markboth{IEEE WIRELESS COMMUNICATION LETTERS}
%\markboth{Journal of \LaTeX\ Class Files,~Vol.~6, No.~1, January~2007}%
%{R.O Adeogun \MakeLowercase{\textit{et al.}}: An Asymptotic Bound on Estimation and Prediction of MIMO-OFDM Wireless Channels}
\vspace{-5pt}

\maketitle

\begin{abstract}
 In this paper, we derive an asymptotic closed--form expression for the error bound on  extrapolation of doubly selective mobile MIMO wireless channels. 
 The bound shows the relationship between the prediction error and system design parameters such as bandwidth, number of antenna elements, and 
 number of frequency and temporal pilots, thereby providing useful insights into the effects of these parameters on prediction performance. 
 Numerical simulations show that the asymptotic bound provides a good approximation to previously derived bounds while eliminating the need 
 for repeated computation and dependence on channel parameters such as angles of arrival and departure, delays and Doppler shifts.
\end{abstract}


\IEEEpeerreviewmaketitle

\section{Introduction}

\IEEEPARstart{T}{he} development of algorithms for the prediction of MIMO--OFDM channels \cite{Jersy2013,ChangCodebook11,Khwart2009,Changkee2007,Adeogun2014ICC} to mitigate performance degradation resulting from feedback delays in adaptive and limited feedback MIMO-OFDM systems have received considerable attention in recent times. In the design of these algorithms, the ability to compute the lower bound on the estimation and prediction error performance as a function of the channel and system parameters is essential in order to make appropriate design decisions. Moreover, these bounds serve as a basis upon which the performance of the different algorithms can be compared. However, there exist no closed--form expressions relating MIMO--OFDM channel estimation, interpolation and prediction performance to predictor design parameters such as number of antennas, number of samples in the observation segment, number of pilot subcarriers, number of paths and SNR.

In \cite{Wong08}, closed--form expressions for the prediction error in SISO--OFDM channels were derived. Bounds on the interpolation of MIMO--OFDM channels were derived in \cite{Larsen2006} using a vector formulation of the Cramer--Rao bound for a 
function of parameters. Similar bounds for estimation and prediction were proposed in \cite{Larsen2008,Larsen2009}. Although these bounds are useful 
in their own way, their expressions are not easily interpretable. Moreover, their dependence on channel parameters necessitates averaging over 
several realizations of the channel resulting in high computational load particularly for large numbers of samples and antenna elements. An asymptotic
expression for the bound on the prediction of narrowband MIMO channels was derived in \cite{Adeogun2014SPL}.

In this contribution, we derive simple, readily interpretable closed--form expressions for the error bound on MIMO--OFDM channel prediction in the 
asymptotic limit of large number of samples and/or antennas. The bounds are applicable to pilot based channel estimation, interpolation and prediction.
The dependence of these bounds on system parameters, but not on channel parameters, enables them to provide useful insight into system design 
considerations.
\section{Channel Model}
We consider a wideband ray--based MIMO channel model defined as\cite[p.~43]{Bruno2013}
\begin{equation}
\label{eq:eq1}
\mathbf{H}(t,\tau)=\sum_{z=1}^Z \alpha_z\mathbf{a}_{\mathrm{r}}(\mu^{\mathrm{r}}_z)\mathbf{a}^T_{\mathrm{t}}(\mu^{\mathrm{t}}_z)e^{j\omega_z t}\delta (\tau-\tau_z)
\end{equation}
where $Z$ is the number of paths, $\alpha_z$ and $\omega_z$ are the complex amplitude and radian Doppler frequency of the $z$th path and $\tau_z$ is
the delay of the $z$th path. $\mathbf{a}_{\mathrm{r}}(\mu^{\mathrm{r}}_z)$ and $\mathbf{a}_{\mathrm{t}}(\mu^{\mathrm{t}}_z)$ are the receive and transmit array response vectors associated with the $z$th path, respectively, while $\mu^{\mathrm{r}}_z$ and $\mu^{\mathrm{t}}_z$ are the angular frequencies associated with the angles of arrival and departure of the $z$th path, respectively. Note that while \eqref{eq:eq1} is valid for all antenna geometries, we will consider a uniform linear array (ULA) such that $\mathbf{a}_{\mathrm{r}}(\mu^{\mathrm{r}}_z)$ is defined as
\begin{equation}
\label{eq:eq2}
\mathbf{a}_{\mathrm{r}}(\mu^{\mathrm{r}}_z)=[1\quad e^{-j\mu^{\mathrm{r}}_z}\quad e^{-j2\mu^{\mathrm{r}}_z}\quad\cdots\quad e^{-j(N-1)\mu^{\mathrm{r}}_z}]^T
\end{equation}
with $\mu^{\mathrm{r}}_z=2\pi\delta_{\mathrm{r}}\sin\theta_z$. $N$ is the number of receive antenna elements, $\delta_{\mathrm{r}}$ is the inter 
element spacing of the receive array and $\theta_z$ is the angle of arrival of the $z$th path. The transmit array response vector, $\mathbf{a}_{\mathrm{t}}(\mu^{\mathrm{t}}_z)$, is analogously 
defined by replacing $N$ with $M$ and $\mu^{\mathrm{r}}_z$ with $\mu^{\mathrm{t}}_z$ in \eqref{eq:eq2}. The frequency response of the channel is 
obtained via the Fourier transform of \eqref{eq:eq1} as\footnote{It should be noted that although the carrier frequency, $f_c$ may be included in 
the delay term as in \cite{Larsen2008}, it is omitted here since it only result in a shift in the phase of each path.}
\begin{equation}
\label{eq:eq3}
\mathbf{H}(t,f)=\sum_{z=1}^Z \alpha_z\mathbf{a}_{\mathrm{r}}(\mu^{\mathrm{r}}_z)\mathbf{a}^T_{\mathrm{t}}(\mu^{\mathrm{t}}_z)e^{j(\omega_z t-2\pi f\tau_z)}
\end{equation}
where $f$ denotes the frequency variable. We assume that channel parameters are stationary over the region of interest and that no two paths 
share the same parameter set $\{\alpha_z,\mu^{\mathrm{r}}_z,\mu^{\mathrm{t}}_z,\omega_z,\tau_z\}$ but two or more paths may share any subset of 
the parameter set. Assuming that the system has perfect sample timing and a proper cyclic extension, the sampled frequency response can be expressed 
as
\begin{equation}
\label{eq:eq3}
\mathbf{H}(p,q)=\sum_{z=1}^Z \alpha_z\mathbf{a}_{\mathrm{r}}(\mu^{\mathrm{r}}_z)\mathbf{a}^T_{\mathrm{t}}(\mu^{\mathrm{t}}_z)e^{j(p\nu_z-q\eta_z)}
\end{equation}
where $p$ and $q$ denote the sample and subcarrier index, respectively. $\nu_z=\Delta t\omega_z$ and $\eta_z=2\pi\Delta f\tau_z$ are the normalized 
Doppler frequency and normalized delay, respectively for symbol period $\Delta t$ and subcarrier spacing $\Delta f$. We assume that there 
are $Q$ equally spaced pilot subcarriers in every OFDM symbol and that $P$ equally spaced pilot symbols are available for the estimation, 
interpolation and/or prediction. Let $U_f = \lceil N_{\mathrm{sc}}/Q\rceil$ and $U_t = \lceil N_{\mathrm{pilot}}/P\rceil$  denote the frequency 
spacing (measured in number of subcarriers) between adjacent pilot subcarrier and temporal spacing (in number of OFDM symbols) between adjacent 
pilot symbols, respectively. $N_{\mathrm{sc}}$ is the total number of used subcarriers and $N_{\mathrm{pilot}}$ is the number of OFDM symbols in 
the training segment. In order to avoid frequency and time domain aliasing, $U_f$ and $U_t$ are chosen such 
that $\Delta f\tau_{\mathrm{max}}U_f\leq 1$ and $2\Delta t\omega_{\mathrm{max}}U_t\leq 1$ \cite{Hooman2008a},
%\begin{equation}
%\Delta f\tau_{\mathrm{max}}U_f\leq 1
%\end{equation}
%and
%\begin{equation}
%2\Delta t\omega_{\mathrm{max}}U_t\leq 1
%\end{equation}
where $\tau_{\mathrm{max}}$ and $\omega_{\mathrm{max}}$ are the maximum path delay and Doppler frequency, respectively. We denote the frequency and time indices of the pilots as $q' = qU_f; \quad q=0,1,2,\cdots,Q-1$ and $p' = pU_t;\quad p=0,1,2,\cdots, P-1$, respectively. We represent entry $(n,m)$ of \eqref{eq:eq3} as
\begin{equation}
\label{eqn1}
h(n,m,p,q)=\sum_{z=1}^Z \alpha_z e^{j(p\nu_z-(n-1)\mu^{\mathrm{r}}_z-(m-1)\mu^{\mathrm{t}}_z-q\eta_z)}
\end{equation}
%for all $n=1,\cdots,N$, $m=1,\cdots,M$ and $p=0,\cdots,P-1$.
We assume that for the purpose of channel estimation, interpolation and/or prediction, $PQ$ samples of the channel frequency response are known 
either from channel estimation or measurement. In practice, the channel estimates contain an error resulting from noise and interference, which 
we model as a summation of the true channel and a noise term \cite{Larsen2006}
\begin{equation}
\label{eq:eq6}
\hat{h}(n,m,p,q)=h(n,m,p,q)+w(n,m,p,q)
\end{equation}
where $w\sim\mathcal{C}\mathcal{N}(0,\sigma^2)$. We will henceforth remove the indices in parenthesis and denote $h(n,m,p,q)$ as $h$.
%Assuming that for convenience, the $Z$ measured samples are collected into an $NMZ\times 1$ vector $\mathbf{h}=[\mathbf{h}^T(1)\quad\cdots\quad\mathbf{h}^T(Z)]^T$ which can be shown using \eqref{eq:eq5} to be
%\begin{equation}
%\label{eq:eq7}
%\mathbf{h}=[\mathbf{h}^T(1)\quad \mathbf{h}^T(2)\quad\cdots\quad\mathbf{h}^T(Z)]^T
%\end{equation}
%It can be easily shown using \eqref{eq:eq5} that
%\begin{equation}
%\label{eq:eq8}
%\mathbf{h}=\sum_{p=1}^P \alpha_p(\mathbf{a}_r(\mu^r_p)\otimes\mathbf{a}_t(\mu^t_p)\otimes\mathbf{a}_d(\nu_p))
%\end{equation}
%where $\mathbf{a}_d(\nu_p)=[1 \quad e^{j\nu_p}\quad e^{j2\nu_p}\quad\cdots\quad e^{j(Z-1)\nu_p}]^T$. A matrix representation of \eqref{eq:eq8} is thus
%\begin{equation}
%\label{eq:eq9}
%\mathbf{a}_d(\nu_p)=[1 \quad e^{j\nu_p}\quad e^{j2\nu_p}\quad\cdots\quad e^{j(Z-1)\nu_p}]^T
%\end{equation}

\section{Asymptotic Error Bound}
We now derive a simple and easily interpretable closed--from expression for the lower bound on prediction mean square error (MSE) in the asymptotic case of large $N$, $M$, $P$ and/or $Q$. We assume that estimation, interpolation or prediction are based on estimation of the parameters of the channel using the available pilot channels followed by estimation, interpolation or prediction for the desired frequency or time location using the estimated parameters. Let the channel parameter vector be denoted as\footnote{Note that although the noise variance $\sigma^2$ can also be included as an element of $\boldsymbol{\Theta}$, it is omitted here since this does not affect the expression for the prediction error bound.}
\begin{equation}
\boldsymbol{\Theta}=\left[\boldsymbol{\theta}_1,\boldsymbol{\theta}_2,\cdots,\boldsymbol{\theta}_Z\right]
\end{equation}
where
\begin{equation}
\boldsymbol{\theta}_z=[\mathfrak{R}(\alpha_z)\quad\mathfrak{I}(\alpha_z)\quad\mu^{\mathrm{r}}_z\quad\mu^{\mathrm{t}}_z\quad\nu_z\quad\eta_z]
\end{equation}
$\mathfrak{R}(\cdot)$ and $\mathfrak{I}(\cdot)$ denote the real and imaginary parts of the associated complex number, respectively. Since our model represents a non-linear function of the channel parameters, the mean square error bound (MSEB) can be found using the Cramer--Rao lower bound (CRLB) for functions of parameters \cite{Kay1}
\begin{equation}
\label{eq:eq14}
\mathrm{MSEB}(p,q)=\sum_{n=1}^N\sum_{m=1}^M \frac{\partial h}{\partial \boldsymbol{\Theta}}[\mathbf{J}(\boldsymbol{\Theta})]^{-1}\frac{\partial h}{\partial \boldsymbol{\Theta}}^H
\end{equation}
where $\mathrm{MSEB}(p,q)=\mathbb{E}[(\mathbf{\hat{h}}(p,q)-\mathbf{h}(p,q))^H(\mathbf{\hat{h}}(p,q)-\mathbf{h}(p,q))]$, $\mathbf{J}^{-1}(\boldsymbol{\Theta})$ is the CRLB on the variance of the channel parameter estimates. The Jacobian in \eqref{eq:eq14} is given by
\begin{align}
 \label{eq:eq15}
 \frac{\partial h}{\partial \boldsymbol{\Theta}}&=\left[\frac{\partial h}{\partial \boldsymbol{\theta}_1}\quad\frac{\partial h}{\partial \boldsymbol{\theta}_2}\quad\cdots\quad\frac{\partial h}{\partial \boldsymbol{\theta}_Z}\right]
\end{align}
$\mathbf{J}(\boldsymbol{\Theta})$ is the Fisher information matrix (FIM), entries of the which can be evaluated element-wise using Bangs formula \cite{Kay1},
\begin{equation}
\label{eq:ex1}
\left[\mathbf{J}(\boldsymbol{\Theta})\right]_{ij}=\operatorname{Tr}\left[\mathbf{C}^{-1}\frac{\partial \mathbf{C}}{\partial \boldsymbol{\Theta}_i}\mathbf{C}^{-1}\frac{\partial \mathbf{C}}{\partial \boldsymbol{\Theta}_j}\right]+2\mathfrak{R}\left[\frac{\partial \mathbf{h}^H}{\partial \boldsymbol{\Theta}_i}\mathbf{C}^{-1}\frac{\partial \mathbf{h}}{\partial \boldsymbol{\Theta}_j}\right]
\end{equation}
where $\mathbf{C}$ is the noise covariance matrix. We assume that the estimation noise is Gaussian such that $\mathbf{C}=\sigma^2\mathbf{I}$, and thus \eqref{eq:ex1} can be reduced to
\begin{equation}
\label{eqn3}
[\mathbf{J}(\boldsymbol{\Theta})]_{ij}=\frac{2}{\sigma^2}\mathfrak{R}\left(\sum_{q=0}^{Q-1}\sum_{p=0}^{P-1}\sum_{n=1}^N\sum_{m=1}^M\frac{\partial h}{\partial \boldsymbol{\Theta}_i}\frac{\partial h}{\partial \boldsymbol{\Theta}_j}^H\right)
\end{equation}
%\begin{figure}[t]
%\centering
%\includegraphics[width=0.8\columnwidth]{PilotArrange.eps}
%\caption{Pilot Arrangement.}
%\label{fig1:fig1}
%\end{figure}
Following straightforward derivation, the partial derivatives with respect to each of the parameters can be shown  to be
\begin{align}
\label{eqn4}
\frac{\partial h}{\partial \mathfrak{R}(\alpha_z)}&=e^{j(p\nu_z-(n-1)\mu^{\mathrm{r}}_z-(m-1)\mu^{\mathrm{t}}_z-q\eta_z)}\\
\frac{\partial h}{\partial \mathfrak{I}(\alpha_z)}&=je^{j(p\nu_z-(n-1)\mu^{\mathrm{r}}_z-(m-1)\mu^{\mathrm{t}}_z-q\eta_z)}\\
\frac{\partial h}{\partial \mu^r_z}&=-j(n-1)\alpha_z e^{j(p\nu_z-(n-1)\mu^{\mathrm{r}}_z-(m-1)\mu^{\mathrm{t}}_z-q\eta_z)}\\
\frac{\partial h}{\partial \mu^t_z}&=-j(m-1)\alpha_z e^{j(p\nu_z-(n-1)\mu^{\mathrm{r}}_z-(m-1)\mu^{\mathrm{t}}_z-q\eta_z)}\\
\frac{\partial h}{\partial \nu_z}&=jpU_t\alpha_z e^{j(p\nu_z-(n-1)\mu^{\mathrm{r}}_z-(m-1)\mu^{\mathrm{t}}_z-q\eta_z)}\\
\frac{\partial h}{\partial \eta_z}&=-jqU_f\alpha_z e^{j(p\nu_z-(n-1)\mu^{\mathrm{r}}_z-(m-1)\mu^{\mathrm{t}}_z-q\eta_z)}
\end{align}
Using \eqref{eqn3} and \eqref{eqn4}--(20) and performing some simplifications, the FIM submatrix corresponding to the $z$th path is obtained as
\begin{equation}
\label{eqn7}
[\mathbf{J}(\boldsymbol{\theta}_z)]=\frac{NMPQ}{\sigma^2}\mathbf{K}
\end{equation}
with
\begin{equation}
\mathbf{K}=\begin{bmatrix}
2 & 0 & 0 & 0 & 0 & 0\\
0 & 2 & 0 & 0 & 0 & 0\\
0 & 0 & \frac{2N^2}{3} & \frac{NM}{2} & -\frac{NPU_t}{2} & \frac{NQU_f}{2}\\
0 & 0 & \frac{NM}{2} & \frac{2M^2}{3} & -\frac{MPU_t}{2} & \frac{MQU_f}{2}\\
0 & 0 & -\frac{NPU_t}{2} & -\frac{MPU_t}{2} & \frac{2P^2U_t^2}{3} & -\frac{QPU_tU_f}{2}\\
0 & 0 & \frac{NQU_f}{2} & \frac{MQU_f}{2} & -\frac{QPU_tU_f}{2} & \frac{2Q^2U_f^2}{3}
\end{bmatrix}
\end{equation}
where we have assumed that $P$, $Q$, $N$ and/or $M$ are large\footnote{It should be noted that $P$, $Q$, $N$ and $M$ do not all have to be large. 
We only require $NMPQ$ to be fairly large so that the approximation $NMPQ\mathbb{E}[g]\approx \sum_{i=1}^{NMPQ} g$ holds.}. Similar to \cite{Larsen2008,Larsen2009} 
we assume that the complex amplitude is $\alpha_z\sim\mathcal{C}\mathcal{N}(0,1)$, such that $\mathbb{E}[|\alpha_z|^2]=1$ and $\mathbb{E}[\mathfrak{R}(\alpha_z)]=\mathbb{E}[\mathfrak{I}(\alpha_z)]=0$. Using the structure of \eqref{eqn7}, the inverse of the FIM submatrix is given by
\begin{equation}
\label{eqn9}
[\mathbf{J}(\boldsymbol{\theta}_z)]^{-1}=\frac{\sigma^2}{NMPQ}\mathbf{K}^{-1}
\end{equation}
where $\mathbf{K}^{-1}$ is the inverse of $\mathbf{K}$ given by
\begin{equation}
\mathbf{K}^{-1}=\begin{bmatrix}
\frac{1}{2} & 0 & 0 & 0 & 0 & 0\\
0 & \frac{1}{2} & 0 & 0 & 0 & 0\\
0 & 0 & \frac{60}{13N^2} & \frac{-18}{13MN} & \frac{18}{13NPU_t}& \frac{-18}{13NQU_f}\\
0 & 0 &\frac{-18}{13MN} & \frac{60}{13M^2} & \frac{18}{13MPU_t}& \frac{-18}{13MQU_f}\\
0 & 0 & \frac{18}{13NPU_t} & \frac{18}{13MPU_t} & \frac{60}{13P^2U_t^2} & \frac{18}{13PQU_tU_f}\\
0 & 0 &  \frac{-18}{13NQU_f} & \frac{-18}{13MQU_f} & \frac{18}{13PQU_tU_f} & \frac{60}{13Q^2U_f^2}
\end{bmatrix}
\end{equation}

%\begin{equation}
%\mathbf{J}^{-1}=\begin{bmatrix}
%\frac{1}{2} & 0 & 0 & 0 & 0 & 0\\
%0 & \frac{1}{2} & 0 & 0 & 0 & 0\\
%0 & 0 & \frac{21}{5N^2} & -\frac{9}{5MN} & \frac{9}{5NPU} & -\frac{3}{5NQ^2V}\\
%0 & 0 & -\frac{9}{5MN} & \frac{21}{5M^2} & \frac{9}{5MPU} & -\frac{3}{5MQ^2V}\\
%0 & 0 & \frac{9}{5 NPU} & \frac{9}{5MPU} & \frac{21}{5P^2U^2} & \frac{3}{5PQ^2UV}\\
%0 & 0 & -\frac{3}{5NQ^2V} & -\frac{3}{5MQ^2V} & \frac{3}{5PQ^2UV} & \frac{2}{Q^2V^2}
%\end{bmatrix}
%\end{equation}
Assuming that the scattering sources are uncorrelated, the FIM has a block diagonal structure
\begin{equation}
\label{eqn10}
[\mathbf{J}(\boldsymbol{\Theta})]=\operatorname{blkdiag}[\mathbf{J}(\boldsymbol{\theta}_1)\quad\mathbf{J}(\boldsymbol{\theta}_2)\quad\cdots\quad\mathbf{J}(\boldsymbol{\theta}_Z)]
\end{equation}
The variance of the parameter estimates are therefore bounded by the diagonal entries of \eqref{eqn10}.  Due to the diagonal structure of the FIM and independence of the FIM submatrices on path parameters, the asymptotic mean square error bound (AMSEB) can be written as
\begin{equation}\label{eqn11}
\mathrm{AMSEB}(p,q)=\sum_{n=1}^N\sum_{m=1}^M \frac{\partial h}{\partial \boldsymbol{\Theta}}[\mathbf{J}(\boldsymbol{\Theta})]^{-1}\frac{\partial h}{\partial \boldsymbol{\Theta}}^H
\end{equation}
For our analysis, we define the signal--to--noise ratio (SNR) as\footnote{This definition is necessary in order to allow fair comparison of the bound across channels with different number of paths} $\mathrm{SNR}=Z/\sigma_Z^2$. Thus, at the same SNR, the noise variance for a $Z$-path channel is $\sigma_Z^2=Z\sigma^2$, where $\sigma^2$ is the noise variance for a single path channel. Substituting \eqref{eqn9} into \eqref{eqn11} and performing some simplifications, we obtain
\begin{align}
\label{eqn12}
\mathrm{AMSEB}(p,q)&=\frac{Z^2\sigma^2}{13PQ}\left[44-\frac{36p}{PU_t}+\frac{60p^2}{P^2U_t^2}-\frac{36q}{QU_f}%\right.\nonumber\\
%&\qquad\left.
+\frac{60q^2}{Q^2U_f^2}-\frac{36qp}{P^2U_t^2Q^2U_f^2}\right]
\end{align}
Based on the assumption of normally distributed complex amplitudes, it can be shown that for a $Z$-path channel $\mathbb{E}[||\mathbf{H}||_F^2]=NMZ$ and the asymptotic normalized mean square error bound (ANMSEB) is obtained from \eqref{eqn12} as
\begin{align}
\label{eqn13}
\mathrm{ANMSEB}(p,q)&=\frac{Z\sigma^2}{13NMPQ}\left[44-\frac{36p}{PU_t}+\frac{60p^2}{P^2U_t^2}%\right.\nonumber\\
%&\qquad\left.
-\frac{36q}{QU_f}+\frac{60q^2}{Q^2U_f^2}-\frac{36pq}{QU_fPU_t}\right]
\end{align}
In this form, the ANMSEB provides useful insights on the effects of the number of antennas, number of frequency and time domain pilots, pilot spacing and SNR on the estimation, interpolation and prediction performance. The following observations can be made from \eqref{eqn13}:
\begin{itemize}
\item The subcarriers near the edge of the frequency band are less predictable than those near the centre.
\item The NMSE grows linearly with an increasing noise variance $\sigma^2$ and number of propagation paths $Z$. This is intuitive and agrees 
with previous results that prediction becomes more difficult with increasing number of paths \cite{paul2001}.
\item The NMSE decreases with increasing number of antennas at either or both ends of the link. This is also intuitive since more structure of the channel is revealed by having more antennas.
\item The contribution to the NMSE from the Doppler frequency (see (19),(26)) and delay estimation (see (20), (26)) lead to the $p^2$ and $q^2$ terms, respectively, demonstrating a quadratic increase with prediction horizon and with frequency. This shows the need to accurately estimate the Doppler frequency and path delays for spatial/temporal prediction and frequency domain interpolation, respectively.
\item The contributions from the cross correlation of error terms involving the Doppler frequency lead to the negative linear term in $p$ in (28), thus reducing the ANMSEB. A plausible explanation for this is that improved Doppler frequency estimates can be obtained from joint parameter estimation. A similar term is obtained from cross terms involving the delays.
%\item For regular sampling interval in the observation and prediction segment, the prediction error is independent of the sampling interval but depends on the number of samples. This shows that it is advantageous to sample the channel...........
\end{itemize}

\section{Numerical Simulations}
\begin{figure}[t]
\centering
\includegraphics[width=0.55\columnwidth]{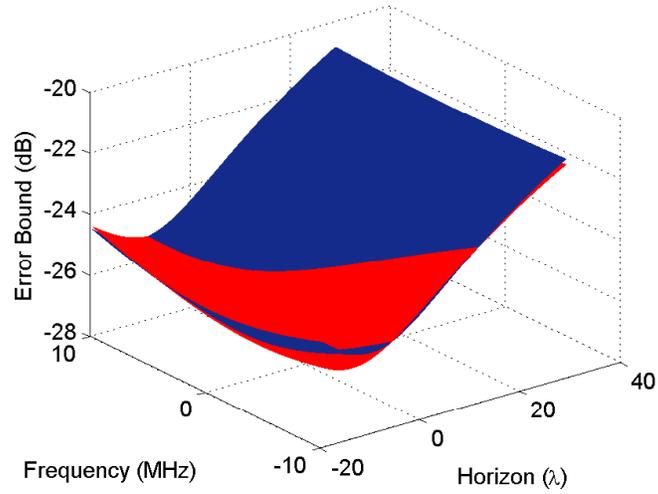}
\caption{Plot of RNMSE versus frequency and horizon ($\lambda$). The upper (blue) surface is the bound in \cite{Larsen2009} and the lower (red) surface is obtained using \eqref{eqn13}.}
\label{fig:fig1}
\vspace{-8pt}
\end{figure}
\begin{figure}[ht]
\centering
\includegraphics[width=0.55\columnwidth]{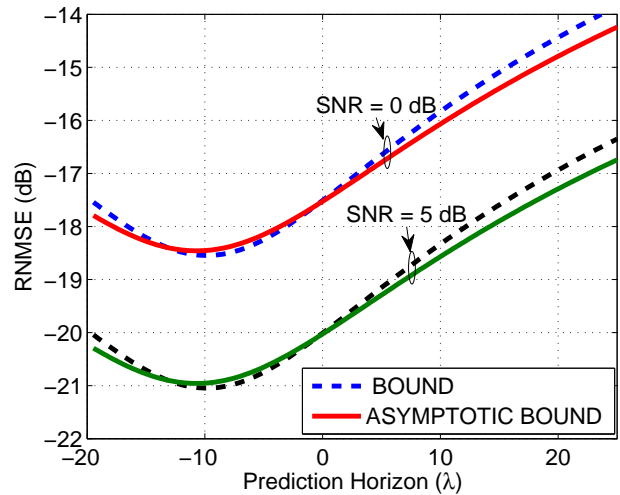}
\caption{Averaged RNMSE versus horizon ($\lambda$)}
\label{fig:fig2}
\end{figure}
%\begin{figure}[h]
%\centering
%\includegraphics[height=0.6\columnwidth,width=1\columnwidth]{WidebandAsymptoticPlot3.eps}
%\caption{Averaged RNMSE versus frequency (MHz)}
%\label{fig:fig3}
%\end{figure}
\begin{figure}[h]
\centering
\includegraphics[width=0.55\columnwidth]{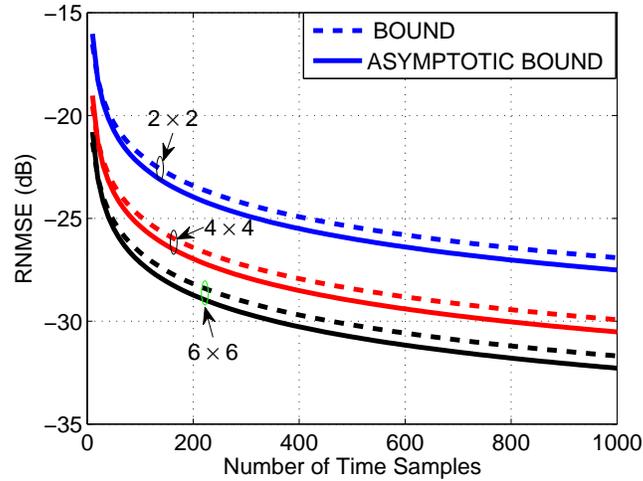}
\caption{Averaged RNMSE versus number of training samples.}
\label{fig:fig4}
\vspace{-8pt}
\end{figure}

\begin{figure}[h]
\centering
\includegraphics[width=0.55\columnwidth]{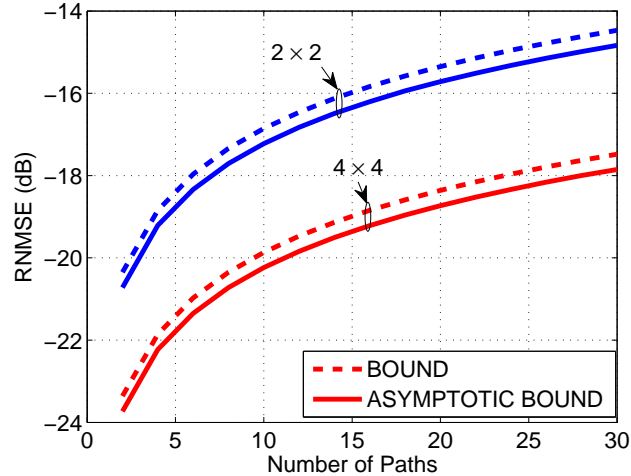}
\caption{Averaged RNMSE versus number of paths.}
\label{fig:fig5}
\end{figure}
In this section, we study the effects of system parameters on the error bounds and compare the asymptotic bound in \eqref{eqn12} with the results in \cite{Larsen2008,Larsen2009}. In order to be consistent with \cite{Larsen2008,Larsen2009}, we consider the root normalized mean square error (RNMSE) defined as $\mathrm{RNMSE}=\sqrt{\mathrm{NMSE}}$. The bound is averaged over 1000 independent channel realizations. We consider a MIMO-OFDM system with bandwidth $B=20\,$MHz, number of subcarriers $N_{\mathrm{sc}}=2048$ and $64$ equally spaced pilot subcarriers. We assume that the channel is sampled at every symbol duration ($U_t=1$). The complex amplitudes are drawn from $\alpha_z \sim \mathcal{C}\mathcal{N}(0,1)$, the angles of arrival and departure are both chosen from a uniform distribution as $\theta^{\mathrm{r}}_z,\theta^{\mathrm{t}}_z\sim \mathcal{U}[-\pi,\pi)$ and the Doppler frequencies are generated from a spatial point of view as $\nu_z=2\pi\Delta x\sin\theta_z^{\mathrm{v}}$, where $\Delta x$ is the spatial sampling interval in wavelengths and $\theta_z^{\mathrm{v}}\sim \mathcal{U}[-\pi,\pi)$ is the angle between the direction of travel of the mobile station and the receive antenna array. The path delays are selected from the delays for the Urban macro (UMA) scenario in the WINNER II/3GPP channel \cite{WINNER2}. We use $\Delta x= 0.2$ for our simulations.

Fig.~\ref{fig:fig1} presents a plot of the asymptotic bound and the bounds in \cite{Larsen2008,Larsen2009} for a two path channel with $P=100$, $Q=64$, $N=2$, $M=2$ and $\mathrm{SNR}=15\,$dB as a function of frequency and horizon (in wavelengths). As seen from the figure, the NMSE bounds increase quadratically in both frequency and temporal horizon and the asymptotic bound approximates the bound very closely.

In Fig.~\ref{fig:fig2}, we plot the RNMSE bounds averaged over frequency versus prediction horizon at $\mathrm{SNR}=[0,5]\,$dB. We observe that over the range considered, the maximum difference between the bounds in \cite{Larsen2008,Larsen2009} and our approximation is only about 0.3\,dB. As expected the bounds increase with horizon but decreases with increasing SNR.%
%A similar plot of the RNMSE bounds versus frequency is presented in Fig.~\ref{fig:fig3}. We observe that the asymptotic bound is close to the actual bound with a maximum difference of about 0.3\,dB.
%Fig.~\ref{fig:fig3} shows the effect of number of antenna at the transmit and receive end of the MIMO link on the prediction NMSE bound using the asymptotic error bound. As can be seen from the figure, the NMSE decreases with increasing number of antennas at either or both ends.

We plot the RNMSE bound versus the number of samples in the observation segment in Fig.~\ref{fig:fig4} for  different numbers of antenna elements at both ends of the link. We observe that, the RNMSE decreases with increasing number of samples. This is intuitively satisfying since an increased number of samples leads to improved parameter estimation and hence to better prediction. It also shows that an increase in the number of transmit and/or receive antenna decreases the RNMSE.

Finally, we show the effects of the number of paths on RNMSE in Fig.~\ref{fig:fig5}. We observe the the RNMSE bounds increases with increasing 
numbers of paths. This agrees with previous observations that propagation channel with dense multipath are more difficult to predict \cite{paul2001}.
\vspace{-5pt}\section{Conclusion}
We have derived simple, easily interpretable and insightful closed--form expressions for the lower bounds on the performance of channel estimation, interpolation and prediction for MIMO--OFDM systems. The bound is obtained using the vector formulation of the Cramer Rao bound for functions of parameters in the asymptotic limits of large frequency and time--domain training samples and number of antennas. The expressions provide useful insights into the effects of system design parameters such as the number of antennas, number of training pilots, noise level, number of paths and pilot spacing on the error performance and are independent of the actual channel parameters. Simulation results show that the asymptotic error bound provides a good approximation to previous formulations while eliminating the need for repeated computation.

\appendix
Consider the expression for the FIM in \eqref{eqn3} and assume that $Q$, $P$, $N$, and/or $M$ are large such that
\begin{equation}
 \label{eqA1}
 \sum_{q=0}^{Q-1}\sum_{p=0}^{P-1}\sum_{n=1}^N\sum_{m=1}^M h\approx QPNM\mathbb{E}[h]
\end{equation}
Using \eqref{eqn3} and \eqref{eqn4}, the diagonal entries of the FIM are obtained as
\begin{align}
 \label{eqA2}
 [\mathbf{J}]_{11} &=[\mathbf{J}]_{22}= \frac{2QPNM}{\sigma^2}\\
 %[\mathfrak{J}]_{22} &= \frac{2QKNM}{\sigma^2}\\
 [\mathbf{J}]_{33} &= \frac{2}{\sigma^2}\left(MPQ(\sum_{n=1}^N (n-1)^2)\mathbb{E}[|\alpha_z|^2]\right)\\
 [\mathbf{J}]_{44} &= \frac{2}{\sigma^2}\left(NPQ(\sum_{m=1}^M (m-1)^2)\mathbb{E}[|\alpha_z|^2]\right)\\
 [\mathbf{J}]_{55} &= \frac{2}{\sigma^2}\left(NPQ(\sum_{k=0}^{P-1} (pU_t)^2)\mathbb{E}[|\alpha_z|^2]\right)\\
 [\mathbf{J}]_{66} &= \frac{2}{\sigma^2}\left(NPK(\sum_{q=0}^{Q-1} (qU_f)^2)\mathbb{E}[|\alpha_z|^2]\right)
\end{align}
Using the identity
\begin{equation}
 \label{eqA3}
 \sum_{a=1}^A a^2 = \frac{A(A+1)(2A+1)}{6}
\end{equation}
and our assumption that the complex amplitude is $\alpha_z\sim\mathcal{CN}(0,1)$, \eqref{eqA2} becomes
\begin{align}
 \label{eqA4}
  [\mathbf{J}]_{33} &= \frac{2}{\sigma^2}\left(\frac{MPQN(N-1)(2N-1)}{6}\right)\nonumber\\
 [\mathbf{J}]_{44} &= \frac{2}{\sigma^2}\left(\frac{NPQM(M-1)(2M-1)}{6}\right)\nonumber\\
 [\mathbf{J}]_{55} &= \frac{2}{\sigma^2}\left(\frac{NMQP(P-1)(2P-1)U_t^2}{6}\right)\nonumber\\
 [\mathbf{J}]_{66} &= \frac{2}{\sigma^2}\left(\frac{NMPQ(Q-1)(2Q-1)U_f^2}{6}\right)
\end{align}
Since $N,M,Q,P>1$, the approximations $A-1\approx A$ and $2A-1\approx 2A$ can be used to simplify 
\eqref{eqA4} as
\begin{align}
 \label{eqA5}
  [\mathfrak{J}]_{33} &= \frac{NMPQ}{\sigma^2}\left(\frac{2N^2}{3}\right)
 &\qquad[\mathfrak{J}]_{44} &= \frac{NMPQ}{\sigma^2}\left(\frac{2M^2}{3}\right)\nonumber\\
 [\mathfrak{J}]_{55} &= \frac{NMPQ}{\sigma^2}\left(\frac{2P^2U_t^2}{3}\right)&\qquad
 [\mathfrak{J}]_{66} &= \frac{NMPQ}{\sigma^2}\left(\frac{2Q^2U_f^2}{3}\right)
\end{align}
The off-diagonal entries of the FIM are obtained following the same procedure.
% or


%\appendices

%\section{}

%\section*{Acknowledgment}

\ifCLASSOPTIONcaptionsoff
  \newpage
\fi



\balance
\bibliographystyle{IEEEtran}
%\small
\bibliography{sample1}
\end{document}